\documentclass[aps,prl,twocolumn,showpacs]{revtex4}
\usepackage{graphicx}
\usepackage{float}

\begin{document}


\title{Probing Majorana neutrinos in the regime of the normal mass hierarchy}



\author{Steven D. Biller \\ {\it Department of Physics, University of Oxford, Oxford OX1 3RH, UK} }



\begin{abstract}
An approach to developing a feasible neutrinoless double beta decay experiment capable of probing Majorana masses in the regime of the nondegenerate normal neutrino mass hierarchy is proposed. For such an experiment, this study suggests that $^{130}$Te is likely the best choice of candidate isotope and that metal-loaded liquid scintillator likely represents the best choice of detector technology. An evaluation of the required loading, scintillator properties and detector configuration is presented. While further development of Te-loaded liquid scintillator is required, recent progress in this area suggests that this task may not be insurmountable. This could open the door for a future experiment of unparalleled sensitivity that might be accommodated in a volume of the order of 10-20 kilotons. To the best of our knowledge, this is the first time that a potentially practical experimental approach to exploring Majorana neutrino masses in the nondegenerate normal hierarchy has been suggested.
\end{abstract}


\maketitle


  The question of whether neutrinos are Majorana in nature is central to modern particle physics. It lies at the heart of the see-saw mechanism that attempts to explain the neutrino mass scale and to fundamental models of Grand Unification and leptogenesis \cite{theory}. Given the scale of neutrino masses, the only known experimental technique that can be practically used to address this question is that of neutrinoless double beta decay ($0\nu \beta \beta$). 
  
  For the standard Majorana mechanism, the rate of $0\nu \beta \beta$ depends on the square of the effective Majorana mass for this process. Due to the mixing of neutrino states, this effective mass can be expressed as $\langle m_{\beta\beta} \rangle = |\sum U^2_{ei}m_i |$, where the sum is over all neutrino mass eigenvalues and $U_{ei}$ are elements of the Pontecorvo-Maki-Nakagawa-Sakata (PMNS) mixing matrix. Observations of neutrino oscillation imply values for both the magnitude of the PMNS elements and the differences between the mass eigenvalues relevant for the three active neutrino species, while the absolute mass scale is experimentally constrained to be $\sim$sub-eV by analyses of tritium beta decay and the cosmic microwave background radiation \cite{constraints}. Furthermore, the matter-enhanced resonant conversion of neutrino flavors in the sun as observed in solar neutrinos \cite{SNO}, when taken together with terrestrial oscillation results \cite{Kam}, constrains the ordering of the neutrino mass states to either be $m_3 > m_2 > m_1$ (``normal" hierarchy) or $m_2 > m_1 > m_3$ (``inverted" hierarchy). For the latter case, this implies an effective Majorana mass in the range of $\sim15-50$meV (unless the mass values are degenerate) whereas, for the normal hierarchy, the mass scale could be lower by an order of magnitude or more.

The difficulty of searching for such a rare decay in an energy region where potential backgrounds are numerous and the parameter of interest only scales with the square root of the lifetime sensitivity makes this a particularly challenging experimental area. Furthermore, theoretical uncertainties in the nuclear matrix elements that dictate expected $0\nu \beta \beta$ rates for a given isotope are large (roughly a factor of 4 in rate for a given effective Majorana mass). Accordingly, progress has been slow and the next generation of instruments is only beginning to approach sensitivity levels relevant to the inverted hierarchy. The notion of a practical experiment that could achieve sensitivities to Majorana masses in the nondegenerate normal hierarchy has seemed fanciful. The purpose of this paper is to seriously assess this possibility based on modest extrapolations from our current understanding of isotope availability, matrix elements and experimental techniques in order to elucidate a realistic path of research and development that might lead to such a measurement.

\section{Choice of Isotope}

To provide initial guidance, some order-of-magnitude comparisons of candidate isotopes are given in Table 1, including approximate quantities and costs of material that would be required to achieve a rate of one $0\nu \beta \beta$ decay per year for an effective Majorana neutrino mass of 2.5meV ($\sim$center of the normal hierarchy phase space as the eigenvalue of the smallest neutrino mass eigenstate approaches zero). The expected signal from isotopes with Q values less than 2 MeV are predicted to have substantially lower rates as well as being much more susceptible to radioactive backgrounds, and so are excluded from the table. For the purposes of this study, a financially viable experiment will be defined as having a total cost of less than one billion U.S. dollars. 

It is conceivable that new, less expensive methods of isotope enrichment might be developed. While there is considerable room for speculation here, an indication of what might be achieved in the best case can be drawn from the fact that the easiest and least expensive of the isotopes to separate at present is $^{136}$Xe, which can directly undergo gas centrifugation. However,  the current cost still amounts to $\sim$\$20 per gram of isotope \cite{Xe}. Thus, even if it were assumed that similar costs might somehow

\newpage

\begin{widetext}

\begin{center}
TABLE 1: Properties of candidate $0\nu\beta\beta$ isotopes. 
\end{center}

\begin{center}
\begin{footnotesize}
\begin{tabular}{|c|c|c|c|c|c|c|c|c|c|c|c|c|} \hline
              &  Q      & percent & element  &   $G^{0\nu}$           &    $M^{0\nu}$   & $T^{0\nu}_{1/2}$ for                    & tons of   & equivalent   &annual world & natural  & enriched &$0\nu / 2\nu$\\
Isotope & (MeV) & natural & cost \cite{Commodity}& ($10^{-14}$/yr) &  (avg) & 2.5meV & isotope for & natural & production \cite{Commodity} &elem. cost   & at \$20/g &rate \cite{constraints}\cite{2nu} \\
              &            & abund.  & (\$/kg) &   \cite{Kotila}      &  \cite{Dueck} &     ($10^{29}$yrs)                        & 1 ev/yr  & tons  & (tons/yr) & (\$M) & (\$M) & ($10^{-8}$)    \\
\hline

$^{48}$Ca & 4.27 & 0.19 & 0.16 & 6.06 & 1.6   & 2.70 & 31.1    & 16380 & 2.4$\times10^8$ &2.6    & 622 & 0.016 \\
$^{76}$Ge & 2.04 & 7.8   & 1650 & 0.57   & 4.8 & 3.18 & 58.2 & 746   &   118    & 1221 & 1164 & 0.55 \\
$^{82}$Se & 3.00 & 9.2   & 174      & 2.48   & 4.0 & 1.05 & 20.8 & 225  &    2000        & 39       & 416 & 0.092 \\
$^{96}$Zr & 3.35 & 2.8   & 36      & 5.02   & 3.0 & 0.93 & 21.4 & 763     &  1.4$\times10^6$  & 27  & 427 & 0.025 \\
$^{100}$Mo & 3.04 & 9.6   & 35      & 3.89   & 4.6 & 0.51 & 12.2 & 127   &   2.5$\times10^5$  & 4.4  & 244 & 0.014 \\
$^{110}$Pd & 2.00 & 11.8   & 23000   & 1.18   & 6.0 & 0.98 & 26.0 & 221  &       207   & 5078       & 521 & 0.16 \\
$^{116}$Cd & 2.81 & 7.6   & 2.8        & 4.08   & 3.6 & 0.79 & 22.1 & 290    &   2.2$\times10^4$  & 0.81  & 441 & 0.035 \\
$^{124}$Sn & 2.29 & 5.6   & 30         & 2.21   & 3.7 & 1.38 & 41.2 & 736      &  2.5$\times10^5$  & 22   & 825 & 0.072 \\
$^{130}$Te & 2.53 & 34.5   & 360      & 3.47   & 4.0 & 0.75 & 23.6 & 68    &   $\sim$150      & 24       & 471 & 0.92 \\
$^{136}$Xe & 2.46 & 8.9   & 1000      & 3.56   & 2.9 & 1.40 & 45.7 & 513   &       50         & 513       & 914 & 1.51 \\
$^{150}$Nd & 3.37 & 5.6   & 42      & 15.4   & 2.7 & 0.37 & 13.4 & 240      &      $\sim 10^4$      & 11       & 269 & 0.024 \\

\hline
\end{tabular}
\end{footnotesize}
\end{center}
\end{widetext}

\noindent   be achieved for all isotopes, {\em none} of these are likely to be financially viable for the normal hierarchy, particularly given that required isotope levels will almost certainly need to be several times larger than those in the table in order to cope with potential backgrounds. We are therefore left only to consider natural abundances.

Current world production levels and prices appear to rule out the use of $^{76}$Ge, $^{110}$Pd and $^{136}$Xe, even if one allows for the possibility of large excursions from current commodity trends. Furthermore, the extreme quantities of raw material required also make $^{48}$Ca a poor choice.

Of the remaining isotopes, $^{130}$Te has, by far, the largest ratio of expected $0\nu\beta\beta$ to $2\nu\beta\beta$ rates, which considerably relaxes the constraint on the energy resolution required to suppress backgrounds from $2\nu\beta\beta$. This background is particularly important to suppress as it fundamentally scales with the quantity of isotope and is indistinguishable from $0\nu\beta\beta$ near the endpoint. $^{130}$Te also requires the smallest quantity of natural material, offering the prospect of a more practical, compact experiment (which is important for limiting background from  solar neutrinos), if a suitable detector technology can be applied.

\section{Detector Technology}
While there is also scope for speculation regarding the development of new detector concepts and improved background rejection methods, useful guidance can be taken from considering the performance of current experimental methods and their potential extensions. Solid state detectors offer the possibility of excellent energy resolution, but have so far not resulted in experiments with low-enough background levels. For example, the CUORE experiment, which uses TeO$_2$ crystals, has an energy resolution of $\sim$5 keV and a target background level (dominated by radioactivity from the detector surfaces) of 0.01 counts/kg/keV/yr \cite{CUORE}. For a 100-ton instrument, this corresponds to $\sim$10$^4$ events per year in the region of interest. Thus, many orders of magnitude improvement in backgrounds levels would be required, in addition to practical issues associated with scaling up this technology by nearly 3 orders of magnitude. Thin foil tracking approaches, such as SuperNEMO \cite{SuperNEMO}, involve instrumental scales that are proportional to surface area as opposed to volume and, thus, are simply not practical for the hundred-ton scale. Liquid TPC technology, such as that used for EXO \cite{EXO}, might provide an interesting approach if a way could be found to incorporate a large density of tellurium in the TPC medium, but this possibility is considered too speculative for this current study. 

However, one promising technique is that of using loaded liquid scintillator \cite{LiquidScint}. This approach is readily scalable to large isotope masses if reasonable loading levels can be achieved, practical in its simplicity of construction and instrumentation, and has the possibility to achieve very low background rates from external radioactivity using a combination of self-shielding, purification and coincidence-tagging. Furthermore, tellurium has no inherent absorption lines in the wavelength range of bialkali photomultiplier tubes (PMTs), allowing for the possibility of good optical properties. Indeed, the SNO+ Collaboration have already demonstrated the ability to load tellurium at the percent level in scintillator and, accordingly, are now pursuing this as the primary isotope for the $0\nu\beta\beta$ phase of that experiment  \cite{SNO+paper}. 

\section{Scaling Model}

A simplified scaling model for signal and background levels relevant to a Te-loaded liquid scintillation detector will now be used to explore the parameter space of general detector and scintillator characteristics that are required to achieve a given sensitivity. For these purposes, the sensitivity will be defined by the value of $Signal/\sqrt{Background}$ in the context of a simple bin-based analysis, where the locations of the energy bin boundaries are optimized for different $2\nu\beta\beta$ background levels (due to the strong energy dependence of this background). 

The number of signal events can be parameterized as:

\[ S = s M_I T \left(\frac{\langle m \rangle}{2.5 \mbox{meV}}\right)^2 \]

\noindent where $M_I$ is the isotope mass in tons; $T$ is the live time in years; and $s$ is then the number of signal events expected in the energy bin for 1 ton-yr of exposure with an effective Majorana mass of 2.5 meV. Thus, assuming the same parameter values given in Table 1, and for a $\pm$1$\sigma$ bin centered on the endpoint (with $\sigma$ an assumed Gaussian energy resolution), $s$=0.68/23.6 = 0.029. For a bin spanning 0-1.5$\sigma$, this becomes 0.018.

It will be assumed that backgrounds from radioactivity external to the scintillator can be suppressed to negligible levels by an appropriate use of shielding and fiducial volume selection. This leaves $^8$B solar neutrinos as the dominant external background, which is assumed to be flat in energy near the $0\nu\beta\beta$ endpoint and is parameterized as:

\[ B_\odot = b_\odot T M_D \left(\frac{L}{1000 \mbox{pe/MeV}}\right)^{-\frac{1}{2}} \]

\noindent where $M_D$ is the overall fiducial detector mass ({\em i.e.} scintillator plus isotope) in tons; $L$ is the level of scintillation light observed for a given interaction, expressed in terms of the typical number of detected photoelectrons per MeV of deposited energy; and $b_\odot$ is then the expected number of elastic scattering $^8$B solar neutrino interactions within a given energy bin for a 1 ton-year exposure and for a scintillator light level corresponding to 1000 detected pe/MeV. Here it is assumed that the size of the energy bin is defined in units of the energy resolution, which scales as $1/\sqrt{L}$. The expected rate of solar neutrino interactions in the vicinity of the 2.53MeV endpoint is $\sim$0.23/ton-yr/MeV \cite{Jones}. Thus, for a $\pm$1$\sigma$ energy bin centered around the endpoint, $b_\odot = 0.23\times 2 \sqrt{2.53/1000} = 0.023$ (for a bin spanning 0-1.5$\sigma$, this becomes 0.017).

The main radioactive background likely to contaminate the scintillator itself is due to $^{214}$Bi (from the uranium series decay chain). However, it has been shown that organic liquid scintillator can attain levels as low as $10^{-17}$g/g of U/Th \cite{Borexino} and the remaining $^{214}$Bi backgrounds can be very effectively identified and removed via the Bi-Po delayed alpha tag \cite{KamLAND}, rendering it insignificant compared to backgrounds from solar neutrinos. Background from the cosmogenic production of radioactive isotopes due to sea-level exposure of tellurium is a potential concern \cite{IDEA}. However, for the purposes of this study, it will be assumed that such isotopes can be reduced to sufficiently low levels via purification \cite{SNO+paper}. This leaves $2\nu\beta\beta$ decays from the tellurium as the dominant internal background, which is parameterized as:

\[ B_{2\nu} = b_{2\nu} M_I T \left(\frac{L}{1000 \mbox{pe/MeV}}\right)^{-\frac{5.5}{2}} \]

\noindent where the exponent in the last term comes from the observation that a $2\nu\beta\beta$ spectrum convolved with a Gaussian energy resolution, $\sigma\propto1/\sqrt{L}$, results in a rate of events misreconstructed into an energy bin near the endpoint that approximately scales as $\sigma^{5.5}$ (consistent with \cite{Elliott}). $b_{2\nu}$ is then the expected number of $2\nu\beta\beta$ events misreconstructed into the energy bin for a 1 ton-year exposure and a scintillator light level of 1000 pe/MeV. Using the measured $2\nu\beta\beta$ half-life for $^{130}$Te of $7\times 10^{20}$yrs \cite{NEMO3}, for a $\pm$1$\sigma$ bin centered on the endpoint, $b_{2\nu}$=1.47 (for a bin spanning 0-1.5$\sigma$, this becomes 0.148).

Thus, the significance of an observation in terms of the effective number of standard deviations is approximated as $S/\sqrt{B_\odot + B_{2\nu}}$. 

It is conceivable that some portion of the solar neutrino events might be rejected by using Cherenkov radiation to yield directional information relative to the position of the sun. Such a rejection is not out of the question if, for example, the light gathering power of the detector were maximized through the use of high quantum efficiency PMTs and a coverage approaching 100\% (which would also then put the required overall light detection levels within the range of current liquid scintillator formulations). One would then expect to see $\sim$tens of unabsorbed Cherenkov photons. If these could be sufficiently time-separated from the scintillation signal, perhaps by developing a scintillator with a relatively long time constant, then it might be possible to gain enough directional information to reduce the solar neutrino contamination. For backward-directed events (relative to the direction from the sun), the effective number of standard deviations becomes $S/\sqrt{4fB_\odot + 2B_{2\nu}}$, where the factor $f$ accounts for the fraction of solar neutrino events remaining in this sample. The contribution from forward-directed events is therefore $S/\sqrt{4(1-f)B_\odot + 2B_{2\nu}}$, which can be added in quadrature with the former. It is also conceivable that the additional Cherenkov information might  even be used to statistically distinguish 0$\nu\beta\beta$ events from background contamination and test the nature of the mass mechanism via the observed angular distribution of the decay \cite{MassMechanism}, should a large enough signal be observed.

\section{Application to the Normal Hierarchy}

The model developed in the preceding section will now be employed to explore relevant detector-related parameters subject to the requirement of achieving a 90\% C.L. (Baysian) sensitivity to a 2.5meV Majorana neutrino mass with five years of live time, assuming the average value of $M^{0\nu}$ for $^{130}$Te indicated in Table 1.

In Fig. 1, the required isotope mass in tons (the magnitude of which, conveniently, also roughly equals the cost of the corresponding amount of natural tellurium in units of millions of U.S. dollars) is plotted as a function of detector fiducial volume for a variety of scenarios, which are as indicated in the caption. 

\vskip 0.35in

\begin{figure}[H]
\includegraphics[width=87mm]{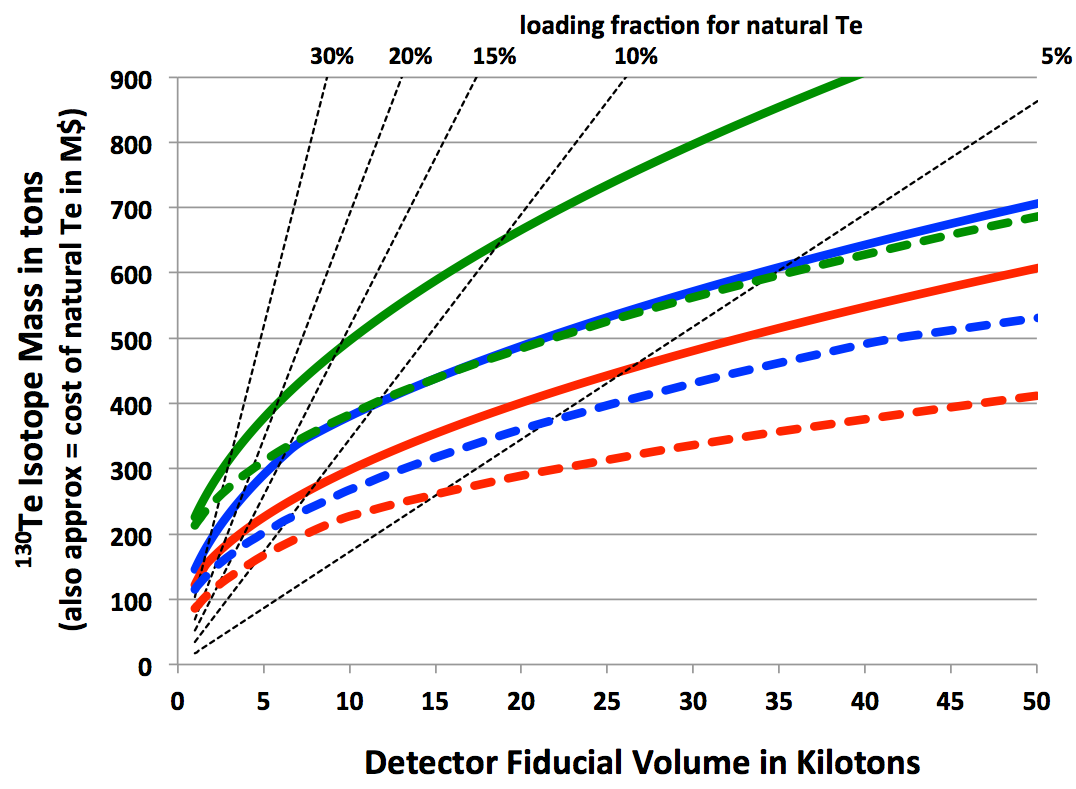}
\caption{Required mass of $^{130}$Te to achieve a 90\%CL sensitivity to a 2.5meV Majorana mass after 5 years of data, assuming $M^{0\nu}$=4. 
Solid curves are for full $^8$B background, whereas long dashes correspond to a 90\% ``forward-backward" directional discrimination of these. Upper curves (green) correspond to a detected scintillation light level of L=1000 pe/MeV; middle curves (blue) to L=1500 pe/MeV; and lower curves (red) to L=2000 pe/MeV. Dotted curves show scintillator loading levels for natural Te.}
\end{figure}

\vskip 0.5in

Thus, for example, the desired sensitivity could be achieved with a 15\% loading of natural Te in a 6kt fiducial volume either with L=1500pe/MeV and full $^8$B backgrounds, or with L=1000pe/MeV and a 90\% directional discrimination of the $^8$B background. Alternatively, this same sensitivity could be achieved with a 10\% loading in a $\sim$11kt volume for L=1000pe/MeV and a 90\% directional discrimination of the $^8$B background. The costs that might be associated with such scenarios would, indeed, appear to be potentially less than a billion U.S. dollars. The required quantities of tellurium would likely be several times the current annual world production rates, but these quantities are still very small compared to estimated reserves \cite{Commodity} and one could imagine a staged deployment where the amount of loading is gradually increased over a period of several years.


\begin{center}
{\bf CONCLUSIONS}
\end{center}

A number of simplifying assumptions have clearly been made in the preceding discussion, not the least of which is the understanding of all potential sources of background and the ability to develop techniques capable of loading Te into scintillator at high-enough levels while maintaining or increasing scintillation light output. However, the purpose of this study has been to attempt a first-order evaluation of basic parameters to assess whether a practical experiment capable of probing Majorana neutrino masses in the regime of the normal hierarchy is feasible in principle and, if so, to indicate a direction for research and development that would be most likely to lead to its realization. The results here suggest that such an experiment may not be beyond the scope of possibility and that the use of Te-loaded liquid scintillator on the scale of $\sim$10-20 kilotons may offer the best (potentially the only) chance for success. Clearly, an important first step would be a demonstration of the basic methodology on a more modest scale. The SNO+ experiment is now pursuing this course, aiming to achieve percent-level loading of Te in a fiducial volume of several hundred tons to test for Majorana neutrinos in the inverted hierarchy.

\begin{acknowledgements}
The author would like to thank the SNO+ Collaboration and, in particular,  Mark Chen and Minfang Yeh for many useful and stimulating discussions. This work has been supported by the Science and Technology Facilities Council of the United Kingdom.

\end{acknowledgements}

\end{document}